\documentclass{aa}
\usepackage{graphicx}
\begin{document}
   \title{Subtracting the photon noise bias from single-mode optical
   interferometer visibilities}

   \author{G. Perrin
          \inst{1}
            }

   \offprints{G. Perrin}

   \institute{LESIA, FRE 2461, Observatoire de Paris, section de 
   Meudon, 5, place Jules Janssen 92190 Meudon, France\\
              \email{guy.perrin@obspm.fr}
              }

   \date{Received 3 July 2002 / Accepted 22 October 2002}

   \abstract{I present in this paper a method to subtract the bias due 
   to source photon noise from visibilities measured with a 
   single-mode optical interferometer. Properties of the processed 
   noise are demonstrated and examples of subtraction on real data 
   are presented.
   }

   \maketitle
%
%________________________________________________________________

\section{Introduction}
The properties of source photon noise are well known.  It follows a 
Poisson distribution whose variance is equal to the average total 
number of photons.  In frequency space, it is a white noise with a 
flat average power spectral density.  In most practical cases where 
the observables are linearly linked to the number of photons detected, 
photon noise can be directly averaged out from the data to reduce its 
variance.  For some applications for which the observables are 
quadratically linked to the number of photons, the data suffer both 
from photon noise and from a bias linked to the variance of the noise.  
This is for example the case in speckle imaging techniques where the 
source spatial intensity distribution is recovered from the power 
spectral density of a short time exposure (\cite{thiebault1994}).  In 
astronomical optical interferometry, the observables are the modulus 
of the visibility and its phase usually expressed as a closure phase 
quantity.  The visibility modulus can be obtained by integrating the 
modulus of the spectrum of interferograms.  However, this estimator is 
biased by the power spectral density of noises as these add to the 
power spectral density of the fringe signal.  An unbiased estimator of 
the modulus of the visibility is obtained by forming the squared 
modulus of the visibility as the power spectral densities of the 
noises can be independently estimated and subtracted. An example in 
interferometry is the computation of the fringe squared visibility in 
the ABCD method where the white light fringe is sampled at four 
$\lambda/4$ spaced optical path differences. An 
unbiased single fringe ABCD estimator is obtained by subtracting the 
source photon noise and the detector noise variances (when the noise 
has a flat spectrum the power spectral density is constant and equal 
to the variance) from the fringe 
power spectral density (\cite{tango1980}).  \\
In single-mode interferometers, beams are spatially filtered by 
single-mode waveguides trading phase fluctuations against intensity 
fluctuations.  A fraction of intensities collected by each aperture 
can be measured to renormalize interferograms to eliminate the 
fluctuations due to turbulence.  The visibility estimator is no longer 
directly linked to the power spectral density of the fringe signal.  I 
demonstrate in the following sections that the classical method 
(explained further in the paper) established by \cite{goodman1985} can 
be rigourously extended to such ratios of physical noisy signals under 
certain assumptions to provide unbiased visibility estimators.  Real 
data reduction cases are presented to illustrate the method.

%__________________________________________________________________

\section{Principles of photon noise bias subtraction}
\begin{figure*}
    \begin{center}
	\vbox{
	\hbox{
	      \includegraphics[width=7.5cm]{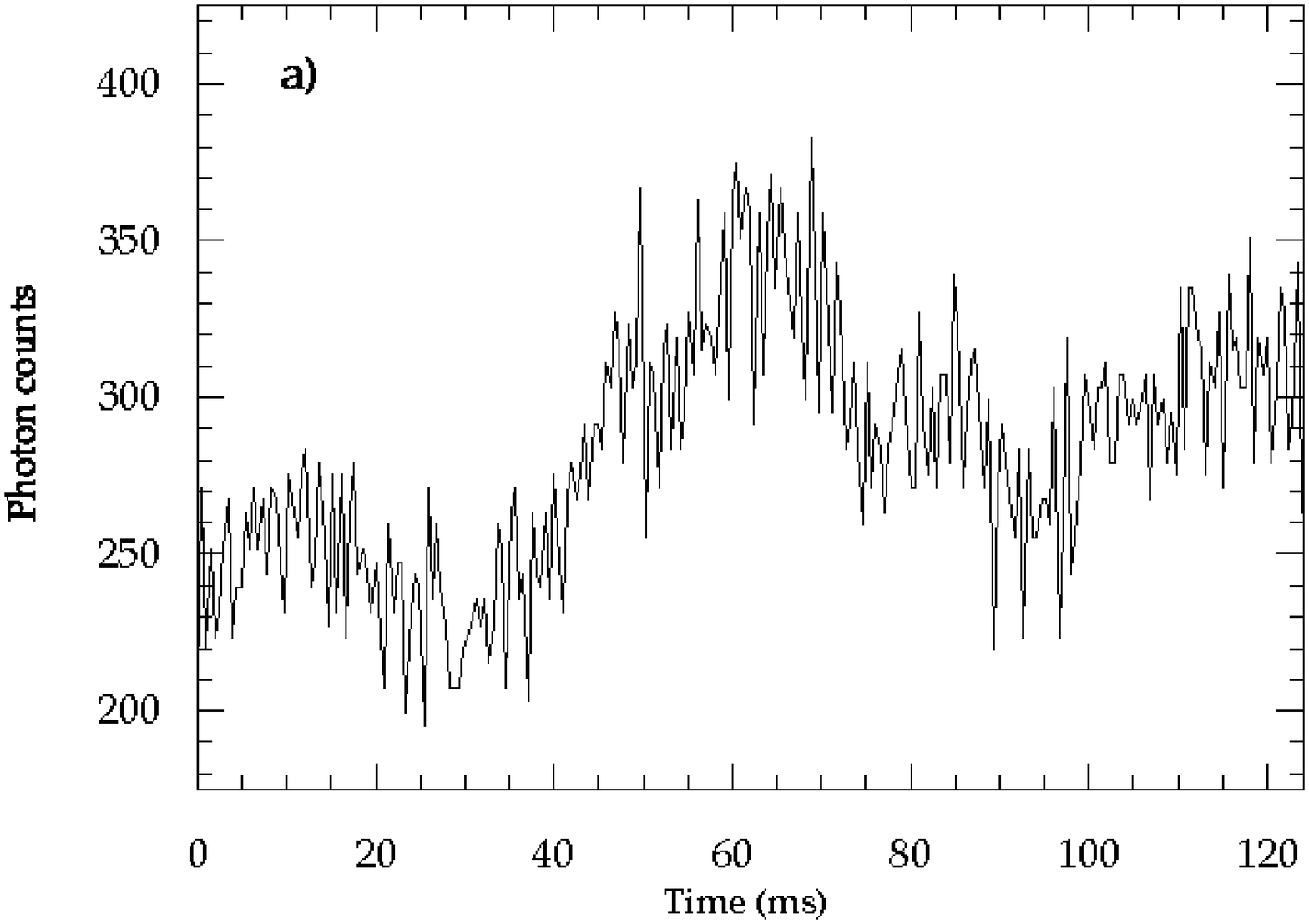}
              \includegraphics[width=7.5cm]{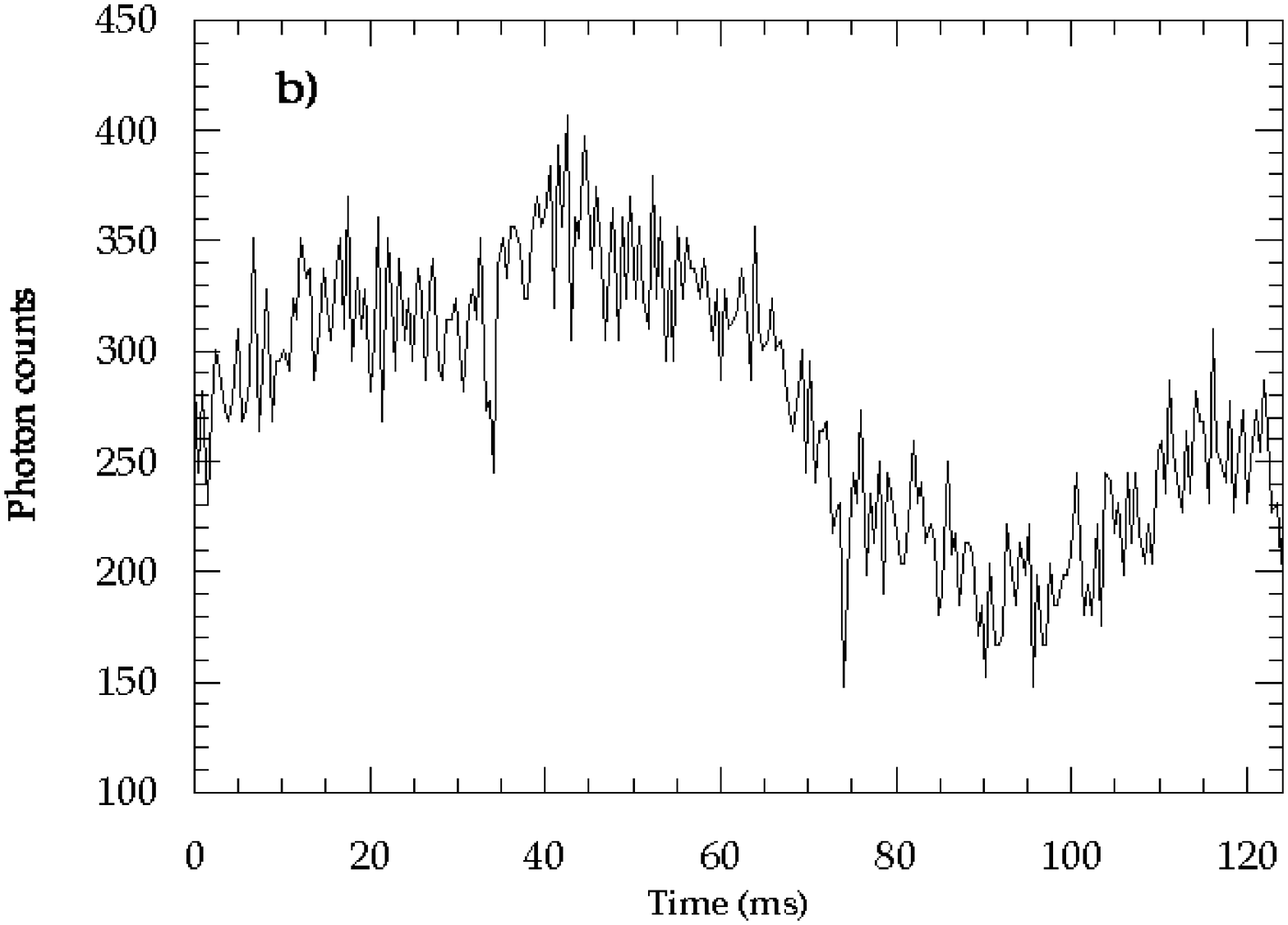}
	     }
	\hbox{
	      \includegraphics[width=7.5cm]{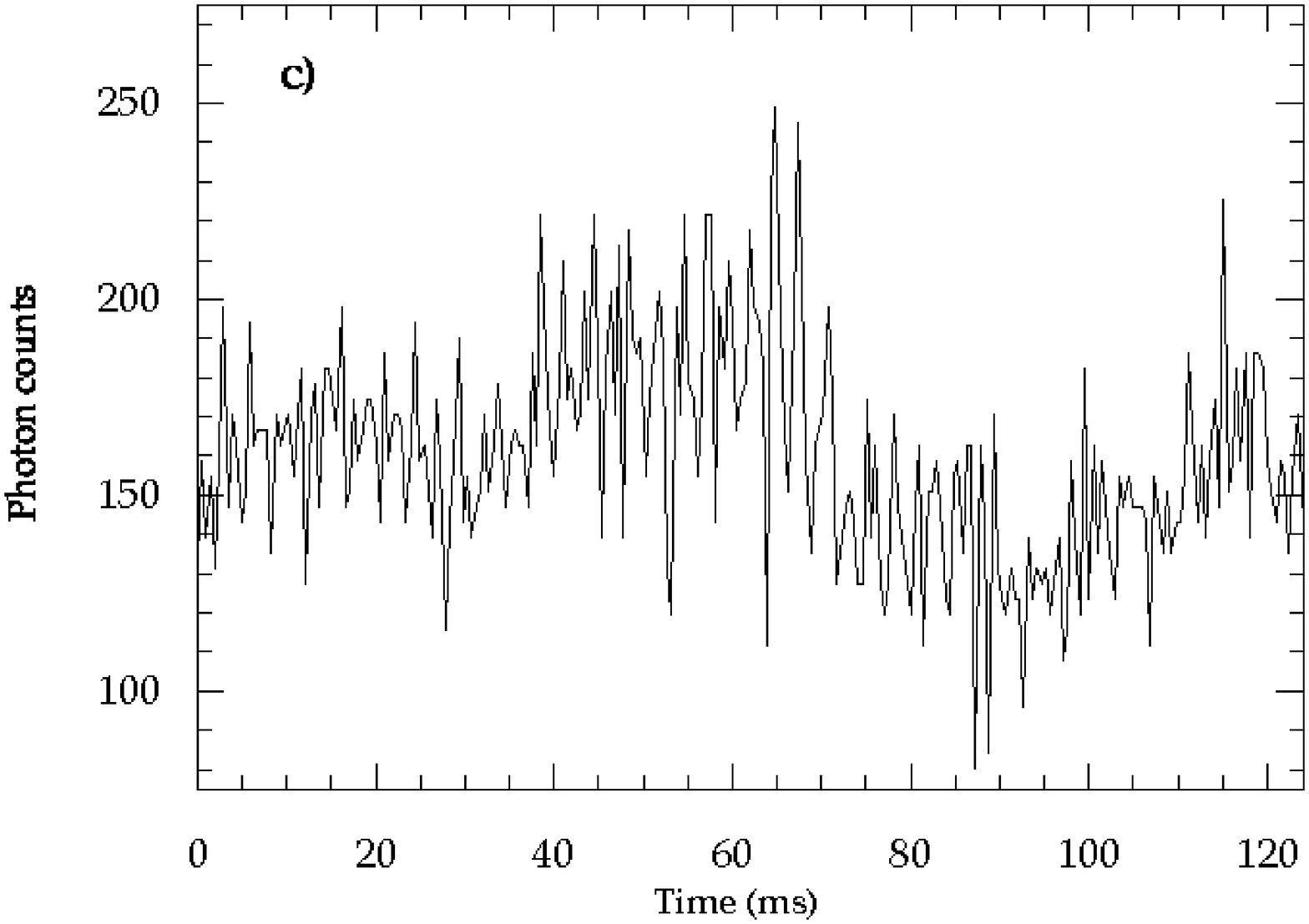}
              \includegraphics[width=7.5cm]{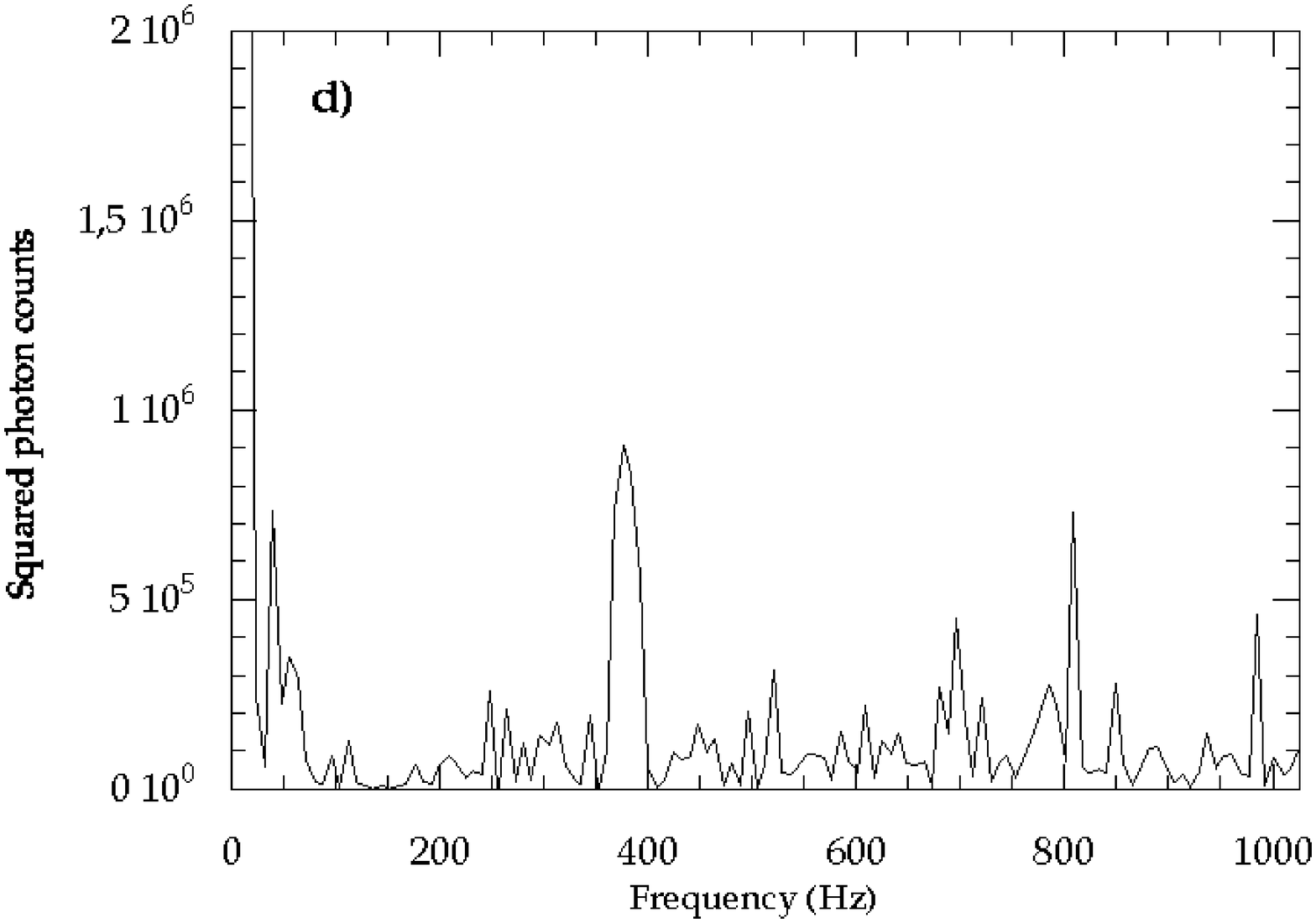}
      } } 
      \caption{Examples of raw signals acquired for Mira in October 
      2000 with FLUOR. a) and b) The two photometric signals 
      $P_{A}(t)$ and $P_{B}(t)$.  c) Interferometric signal 
      $i(t)$.   d) Power spectral density of the interferometric 
      signal.  Energy units are photon counts.  The power spectral 
      density is expressed in squared photon counts.  The fringe peak 
      is located between 350 and 400\,Hz.  Other peaks are due to the 
      noise and disappear when averaging the power spectral densities.  
      The source is resolved hence the fringe amplitude is small and 
      hardly shows up in the interferogram signal.  The fringe peak in 
      the power spectral density is proportional to the squared 
      modulus of the visibility which is obtained by integrating the 
      peak.}
         \label{fig:signauxbruts}
   \end{center}
     \end{figure*}

\subsection{Description of the signals and assumptions}
\begin{figure*}
    \begin{center}
	\vbox{
	\hbox{
	      \includegraphics[width=7.5cm]{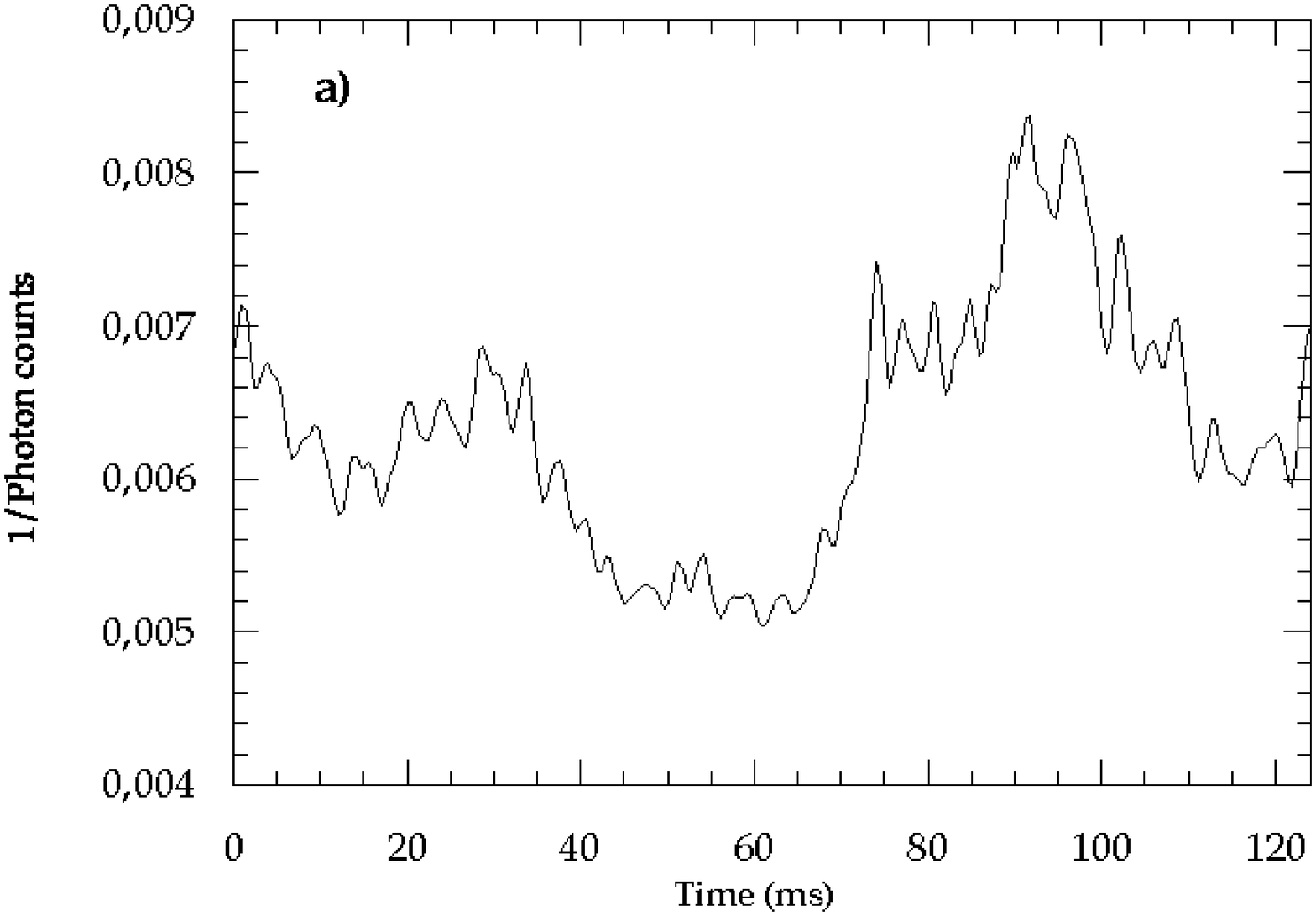}
              \includegraphics[width=7.5cm]{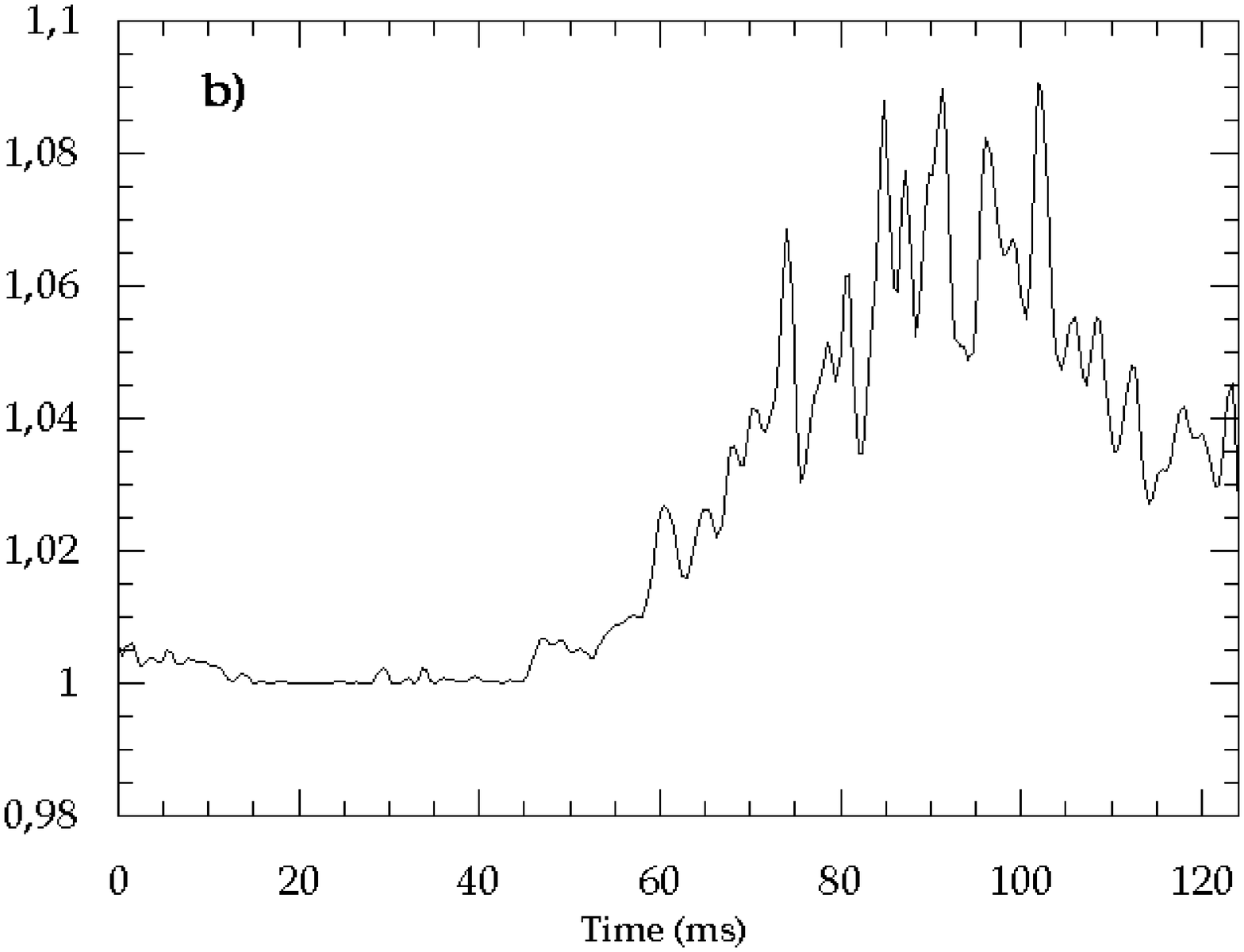}
	     }
	\hbox{
	      \includegraphics[width=7.5cm]{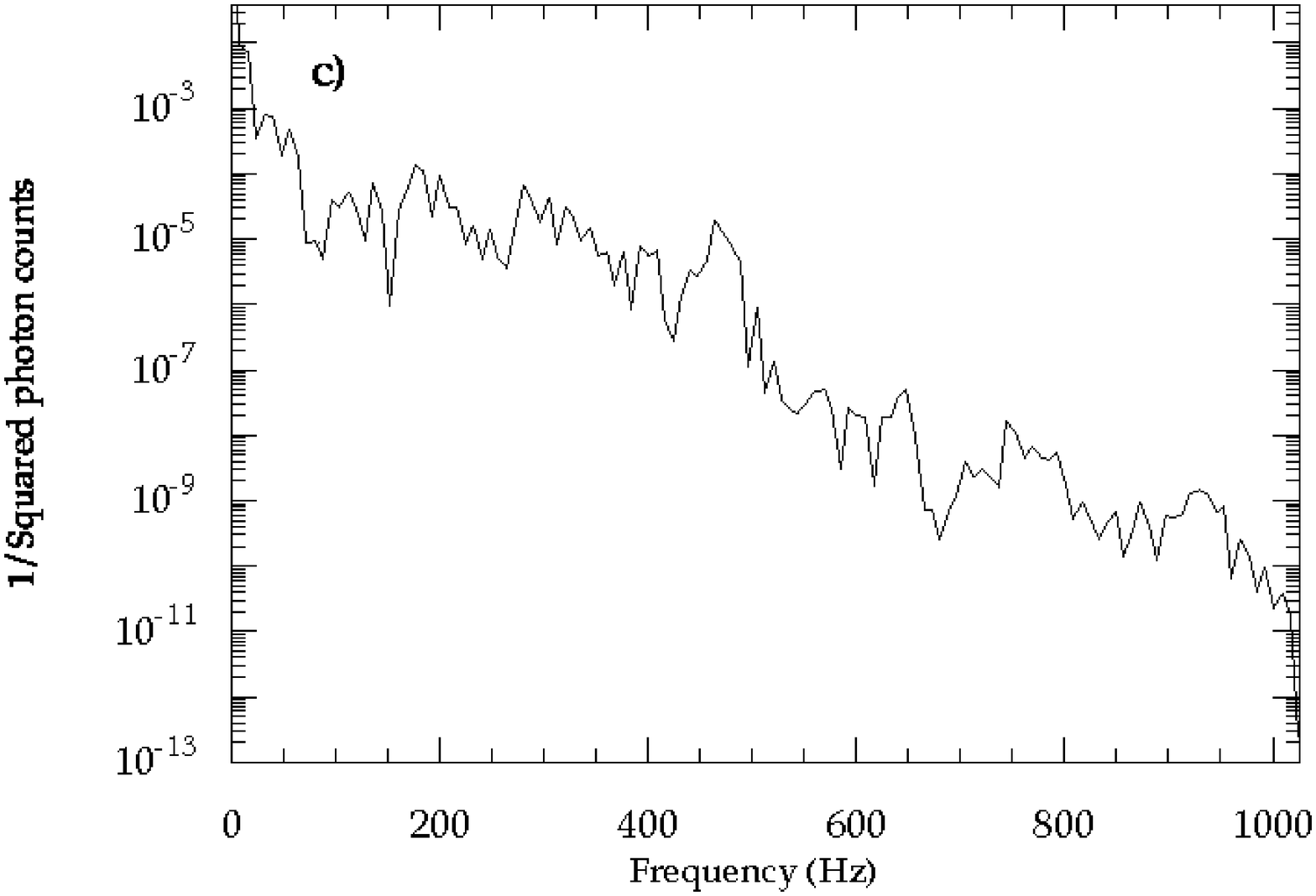}
              \includegraphics[width=7.5cm]{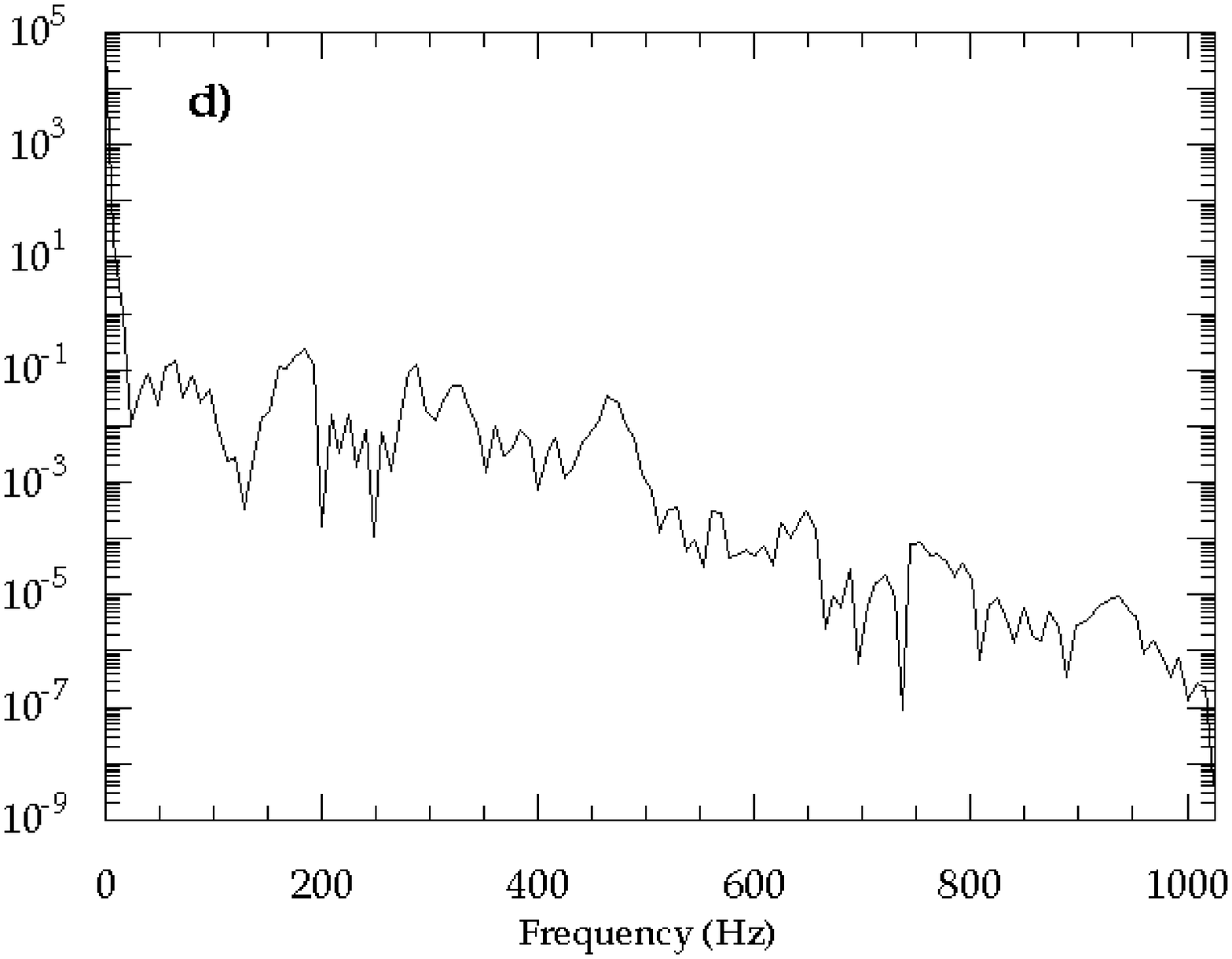}
	}
             }
      \caption{ a) Gain function $g(t)$.  b)
      Continuum function $c(t)$.  c) Power spectral density of 
      the gain function.  d) Power spectral density of the 
      continuum function.  The gain function is homogeneous to the 
      reciprocal of a number of photon counts.  The continuum function 
      has no unit.}
         \label{fig:continugain}
   \end{center}
     \end{figure*}
  
I refer the reader to \cite{foresto1997} for a full description of the 
principle to measure fringe amplitudes (also called coherence factors) 
with a single-mode fiber interferometer and for a detailed description 
of the fringe signal for coaxial interferometers.  Visibility 
calibration will be addressed in a separate paper (\cite{perrin2002}).  
I assume here, for sake of simplicity, that no calibration is required 
and the coherence factor is directly equal to the visibility.  Here I 
will use the more general expression of the interferogram for a 
two-telescope interferometer:
\begin{equation}
    i(x)=P_{A}(x)+P_{B}(x)+2\sqrt{P_{A}(x)P_{B}(x)}\,m(x)
    \label{eq:interferogram}
\end{equation}
where $m$ is an oscillating function containing the fringes.  $P_{A}$ and $P_{B}$ are the 
intensities coupled in the single-mode waveguide at each telescope and 
are called the photometric signals.  $i$ is a function of the optical 
path difference or of time for coaxial beamcombiners.  For multiaxial 
interferometers, $x$ is a focal plane coordinate.  In the following, I 
will deal with coaxial interferometers only and I will use time $t$ as 
the variable.  The method can be easily adapted to multiaxial 
interferometers.  The photometric signals vary in time with turbulence 
and the modulation $m$ varies with turbulence and with the optical 
difference which is varied linearly with time, coding the fringe 
signal in frequency space. $i$ is measured in photon counts and is defined 
as the average signal one would obtain if no noise were present.\\ \\
In the photometric calibration method, the photometric signals need to 
be estimated.  The estimated signals are filtered, the filtering 
function being adjusted to reject most of the noise and keep the 
intensity fluctuations due to turbulence only.  Turbulent fluctuations 
are low frequency fluctuations (limited to frequency ranges of a few 
tens of Hertz).  I note $\overline{P_{A}}$ and $\overline{P_{B}}$ the 
estimated photometric signals suitably low-pass filtered.  The 
important property of the filtered photometric signals is that they 
contain no energy at the fringe frequency and above. \\
I call $g$ the following gain function:
\begin{equation}
    g(t)=\frac{1}{2\sqrt{\overline{P_{A}}(t)\overline{P_{B}}(t)}}
    \label{eq:gain}
\end{equation}
I define the normalized interferogram:

\begin{eqnarray}
    i_{n}(t) & = & g(t).i(t)  \\ \nonumber
             & = &  \frac{P_{A}(t)+P_{B}(t)}{2\sqrt{\overline{P_{A}}(t)\overline{P_{B}}(t)}}
	 +\frac{\sqrt{P_{A}(t)P_{B}(t)}}{\sqrt{\overline{P_{A}}(t)\overline{P_{B}}(t)}}\,m(t)
    \label{eq:In}
\end{eqnarray}
 The first term is mainly a low frequency signal whereas the second 
term is the high frequency signal containing the fringe modulation.
I introduce the continuum function:
\begin{equation}
    c(t)=\frac{\overline{P_{A}}(t)+\overline{P_{B}}(t)}{2\sqrt{\overline{P_{A}}(t)\overline{P_{B}}(t)}}
\end{equation}
 The continuum function is the low frequency part of the normalized 
interferogram.  It is the ratio of the arithmetic and geometric means 
of the photometric signals.  If the two photometric signals are equal 
then the continuum function is equal to 1.  It departs all the more 
from 1 as the photometric beams get unbalanced. The visibility 
estimate is computed from the corrected interferogram:
\begin{eqnarray}
    i_{cor}(t) & = & i_{n}(t)-c(t) \\ \nonumber
               & = &  \frac{\left[P_{A}(t)-\overline{P_{A}}(t)\right]+\left[P_{B}(t)-\overline{P_{B}}(t)\right]}{2\sqrt{\overline{P_{A}}(t)\overline{P_{B}}(t)}} \\ \nonumber
	       &   & + \frac{\sqrt{P_{A}(t)P_{B}(t)}}{\sqrt{\overline{P_{A}}(t)\overline{P_{B}}(t)}}\,m(t)
    \label{eq:Icor}
\end{eqnarray}
\begin{figure*}[t]
    \begin{center}
	\hbox{
	      \includegraphics[width=7.5cm]{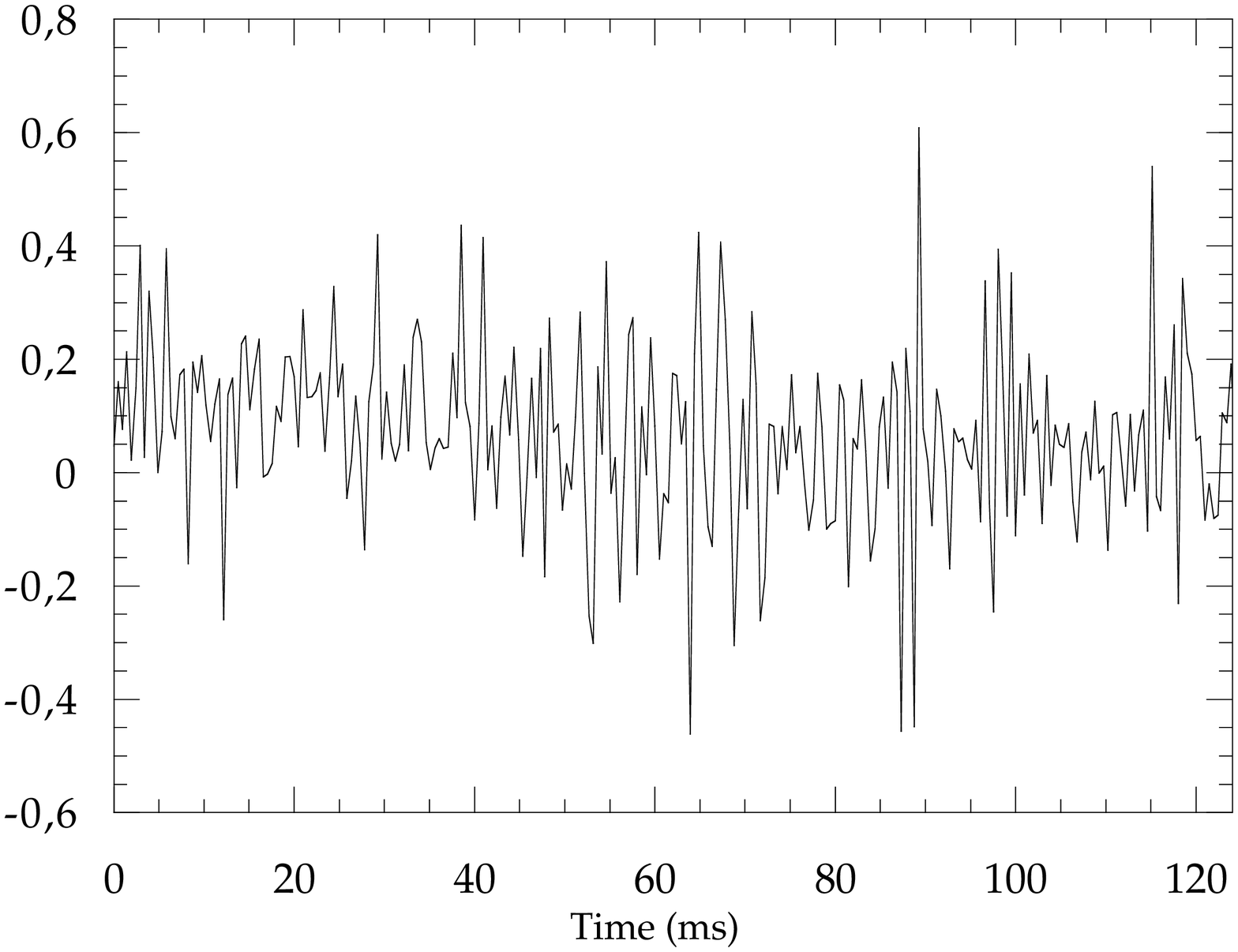}
              \includegraphics[width=7.5cm]{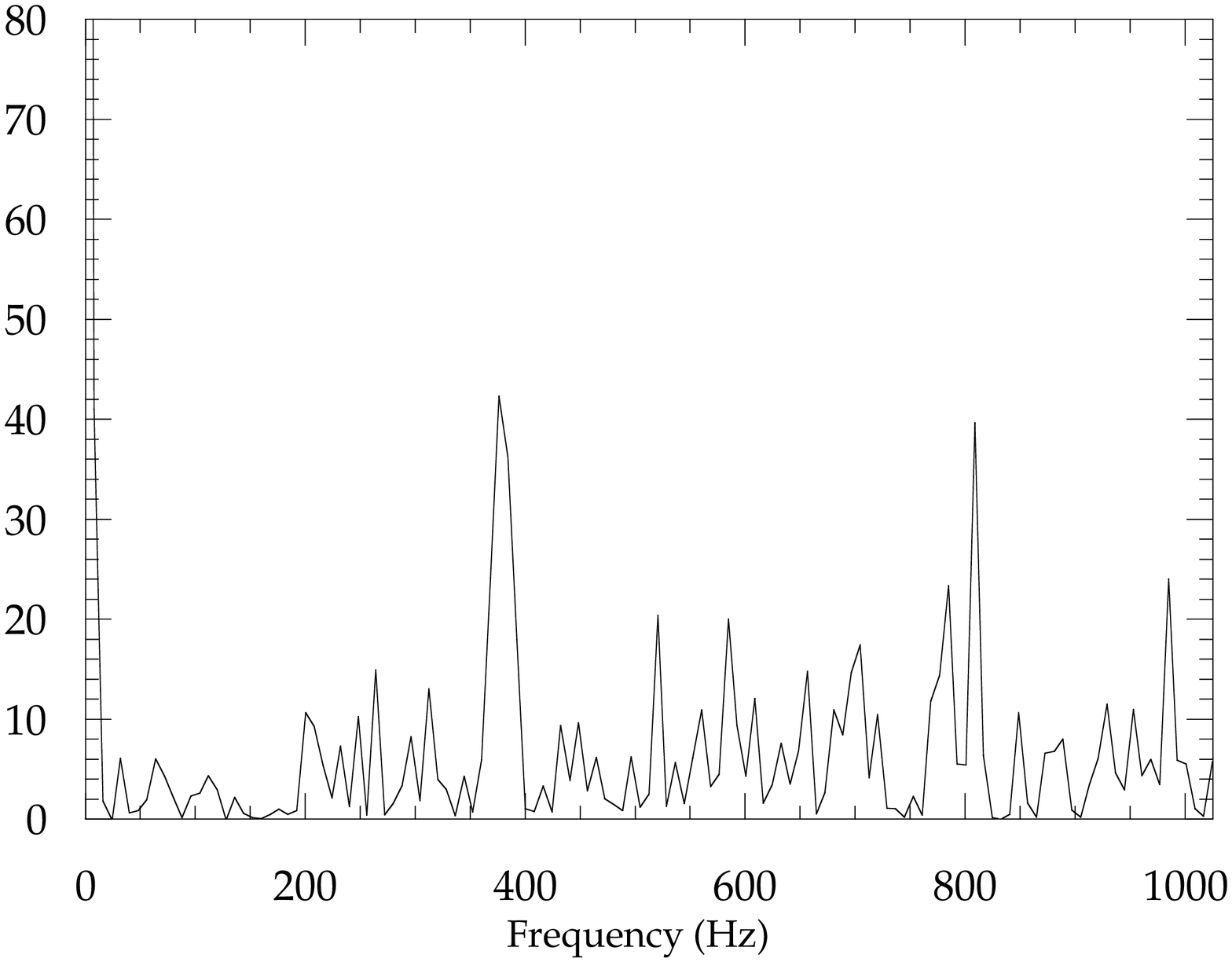}
	     }
      \caption{The corrected interferogram and its power spectral 
      density. The corrected interferogram being the ratio of photon 
      counts has no unit.}
         \label{fig:signaux corriges}
   \end{center}
     \end{figure*}
 In the corrected interferogram the low frequency components due to 
turbulence are eliminated.  If the photometric signals are perfectly 
estimated then the average value of the corrected interferogram is 
equal to zero and the oscillating signal is properly renormalized.  In 
the ideal case where data are noiseless, the corrected interferogram 
is equal to the oscillating function $m$ which is proportional to the 
visibility.  Although the filtered photometric signals do not contain 
energy at the fringes frequency, the gain and continuum function may 
contain some as the residual noise may have been redistributed in the 
frequency domain by the non-linear combinations of the filtered 
photometric signals.  I make the assumption that this high frequency 
noise is negligeable by several orders of magnitude compared to the 
energy of the fringes.  This assumption will be validated in Sect.~3 
on real data.  As a consequence of this assumption, $i_{n}$ and 
$i_{cor}$ share the same photon noise.  This assumption is crucial for 
the success of the method.  Finding an analytical solution to the 
problem of bias subtraction without this assumption is certainly a big 
issue.

\subsection{Method}   
The method to subtract the photon noise bias is a direct 
generalization of that proposed by \cite{goodman1985} and I will use 
the same notations.  In the following only photon noise is considered. 
 Methods to subtract additive noises are well established (see for 
example \cite{foresto1997}). Detector and source photon noises being  
independent the detector noise could be added in the following 
derivation leading to the classical result on power spectral density 
bias by detector noise variance. The primary scope of this paper 
being source photon noise I have decided to keep equations as light  
as possible and not include detector noise in the equations. Detector 
noise will nevertheless be considered in the last section on real data.\\ \\
I call $\tilde{\imath}(t)$ a representation of the interferogram 
in which photon events are represented by Dirac functions.  Assuming 
that the total number of photons in the interferogram is $\tilde{N}$ 
for a given realization, then I can write:

\begin{equation}
    \tilde{\imath}(t)=\sum_{k=1}^{\tilde{N}}\delta(t-t_{k})
\end{equation}
There are two random variables in this expression: the individual 
photon detection times $t_{k}$ and the total number of photons 
$\tilde{N}$ which is equal to $N$ in average.  A model of the 
normalized interferogram can be built:

\begin{equation}       
       \tilde{\imath}_{n}(t)=\sum_{k=1}^{\tilde{N}}g(t)\delta(t-t_{k})=\sum_{k=1}^{\tilde{N}}g(t_{k})\delta(t-t_{k})
\end{equation}
Its Fourier transform is therefore:

\begin{equation}
    \tilde{I}_{n}(f)=\sum_{k=1}^{\tilde{N}}g(t_{k})e^{-2i\pi t_{k}f}
\end{equation}
Thus, the average spectrum of the normalized interferogram is therefore:

\begin{eqnarray}
    \langle \tilde{I}_{n}(f)\rangle &=&\langle \langle \sum_{k=1}^{\tilde{N}}g(t_{k})e^{-2i\pi 
        t_{k}f}\rangle_{t_{k}}\rangle_{\tilde{N}} \\
	&=&\langle \sum_{k=1}^{\tilde{N}}\langle g(t_{k})e^{-2i\pi 
        t_{k}f}\rangle_{t_{k}}\rangle_{\tilde{N}}
\end{eqnarray}
In the above expression, the average on $t_{k}$ does not depend upon 
$k$ and the average normalized interferogram can be written:

\begin{equation}
    \langle \tilde{I}_{n}(f)\rangle=\langle \tilde{N} \langle g(t)e^{-2i\pi tf}\rangle_{t}\rangle_{\tilde{N}}
\end{equation}
The statistics of the photon arrival time $t$ are described by the 
probability density function $\frac{i(t)}{N}$ with $i(t)$ the average 
photon flux.  Hence the average on the arrival times is equal to:
\begin{eqnarray}
\langle g(t)e^{-2i\pi tf}\rangle_{t} & = & 
\int_{-\infty}^{+\infty}g(t)\frac{i(t)}{N}e^{-2i\pi tf}dt \\
& = & \frac{1}{N}I(f) \star G(f) 
\label{eq:conv}
\end{eqnarray}
The $\star$ symbol indicates a convolution. Capital letters are used 
for Fourier transforms.  
Replacing the average on $t$ by this expression, I obtain for the 
average normalized interferogram:

\begin{eqnarray}
    \langle \tilde{I}_{n}(f)\rangle &=&\langle {\tilde{N}} 
    \rangle_{\tilde{N}} \frac{1}{N}G(f) \star I(f) \\
	&=& G(f) \star I(f)
\end{eqnarray}
As a check, the expression of the average interferogram after applying the 
inverse Fourier transform to the above expression is:

\begin{equation}
    \langle \tilde{i}_{n}(t)\rangle = g(t).i(t)
\end{equation}
Let us consider $\tilde{I}_{n}^{(2)}(f)$ the power spectral density of 
the normalized interferogram with $\tilde{I}_{n}^{(2)}(f)= 
|\tilde{I}_{n}(f)|^{2}$.  In the ideal case of non noisy data, the 
integral of the power spectral density is proportional to the squared 
visibility. The integral of the average power spectral density is:

\begin{equation}
    \left\langle \tilde{I}_{n}^{(2)}(f) \right\rangle = 
    {\left\langle {\left\langle {		   
    \sum_{k=1}^{\tilde{N}}\sum_{l=1}^{\tilde{N}} 
    g(t_{k})g(t_{l})e^{-2i \pi (t_{k}-t_{l})f}
} \right\rangle   }_{t_{k},t_{l}} \right\rangle}_{\tilde{N}}
\end{equation}
\begin{figure*}
    \begin{center}
	\vbox{
	\hbox{
	      \includegraphics[width=7.5cm]{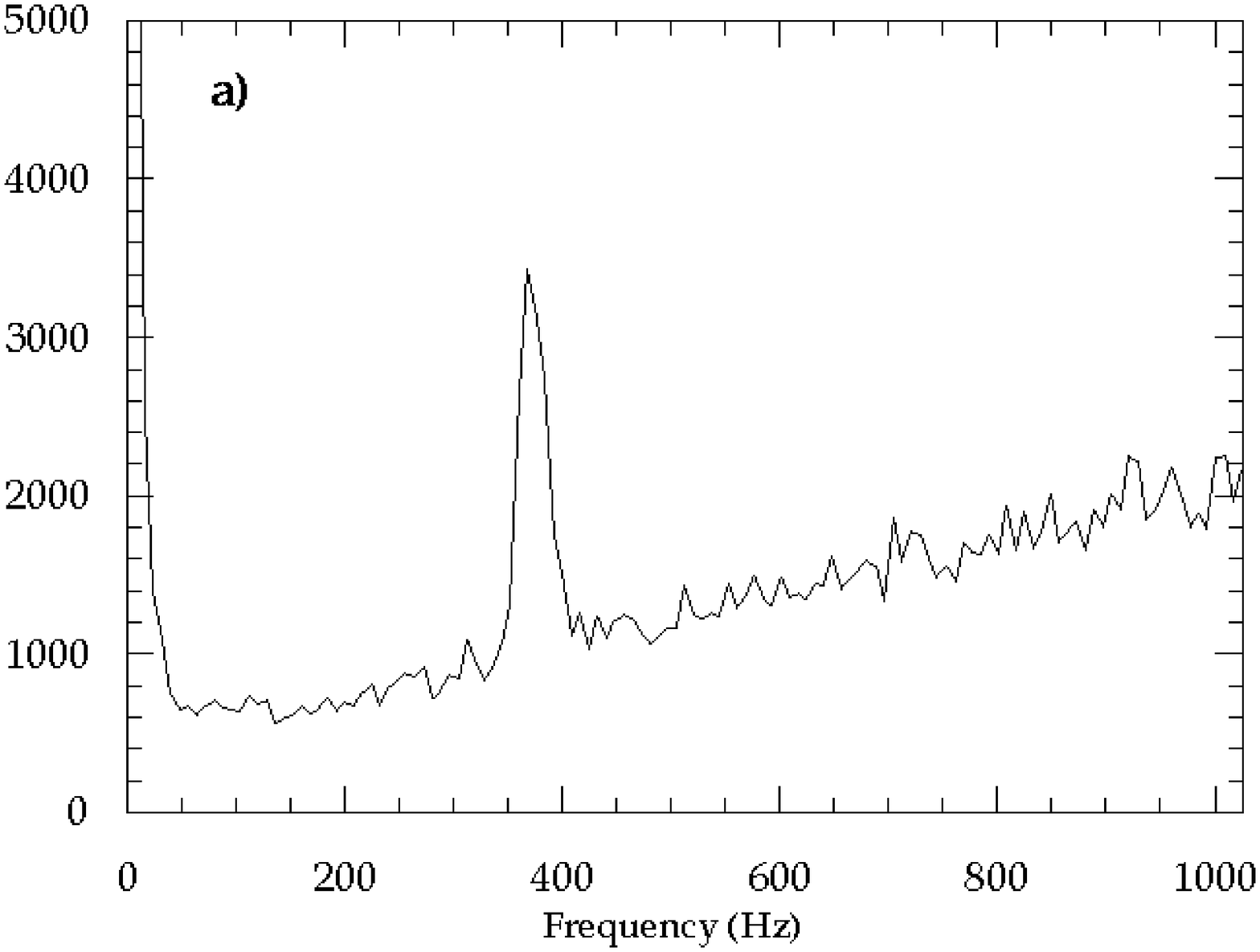}
              \includegraphics[width=7.5cm]{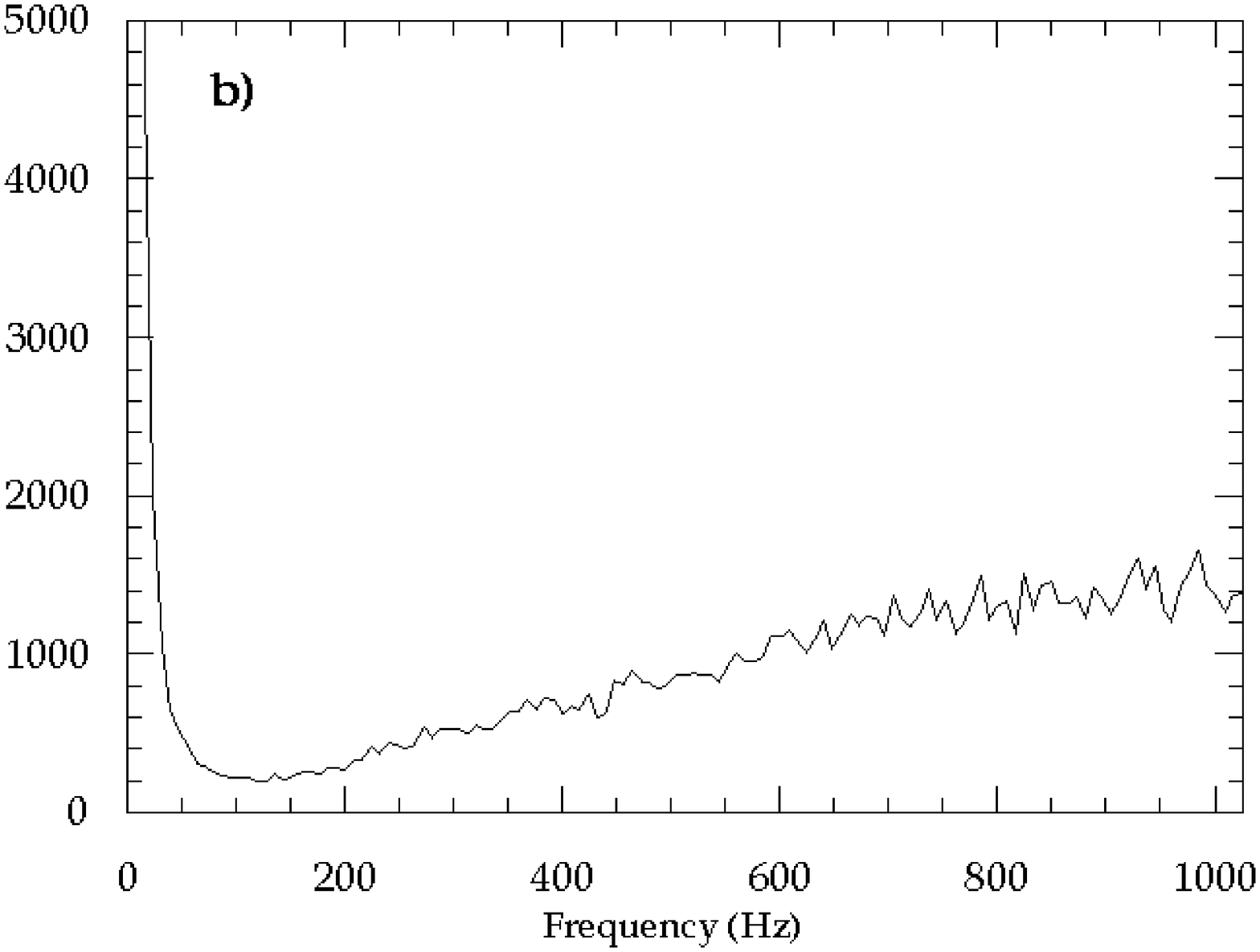}
	     }
	\hbox{
	      \includegraphics[width=7.5cm]{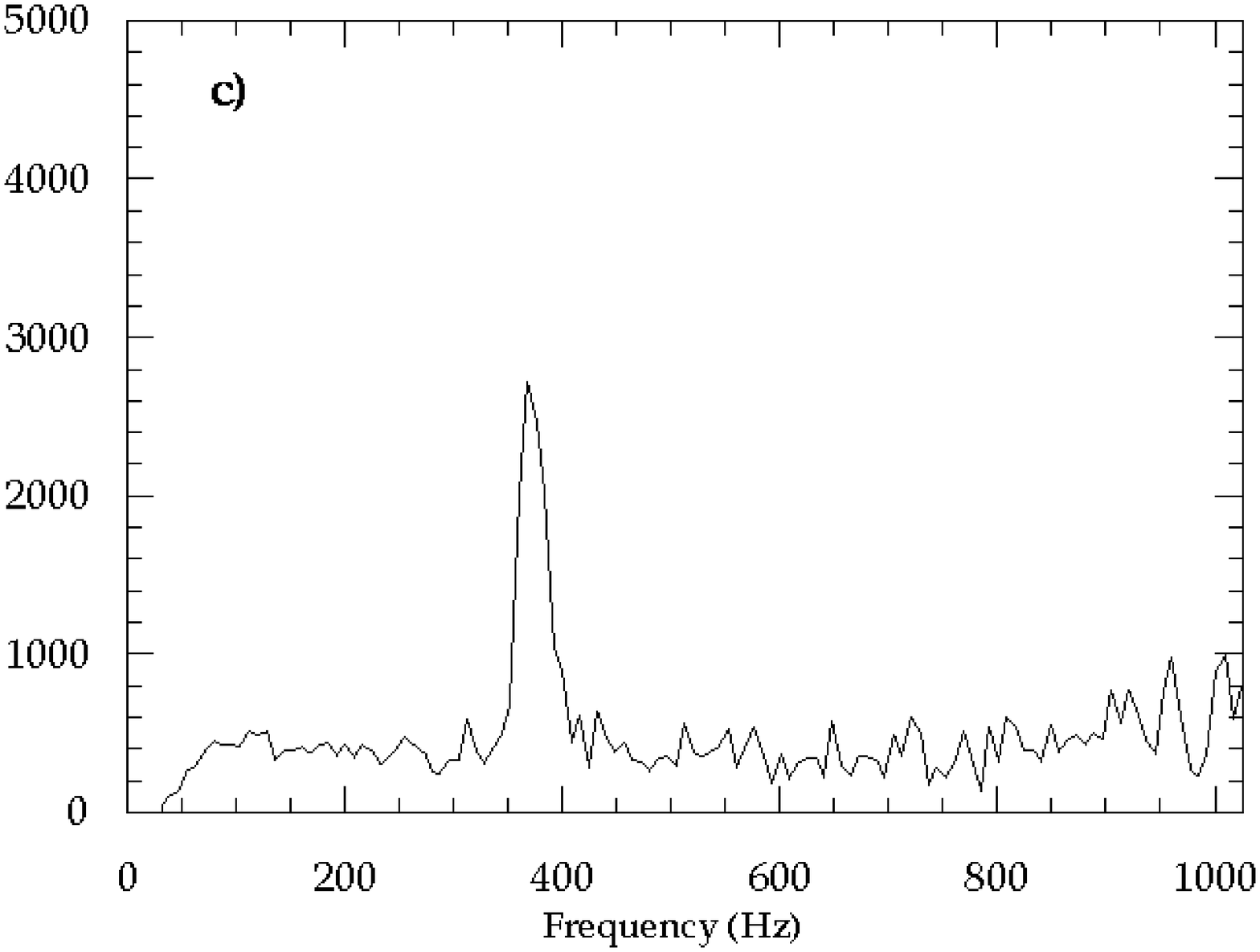}
              \includegraphics[width=7.5cm]{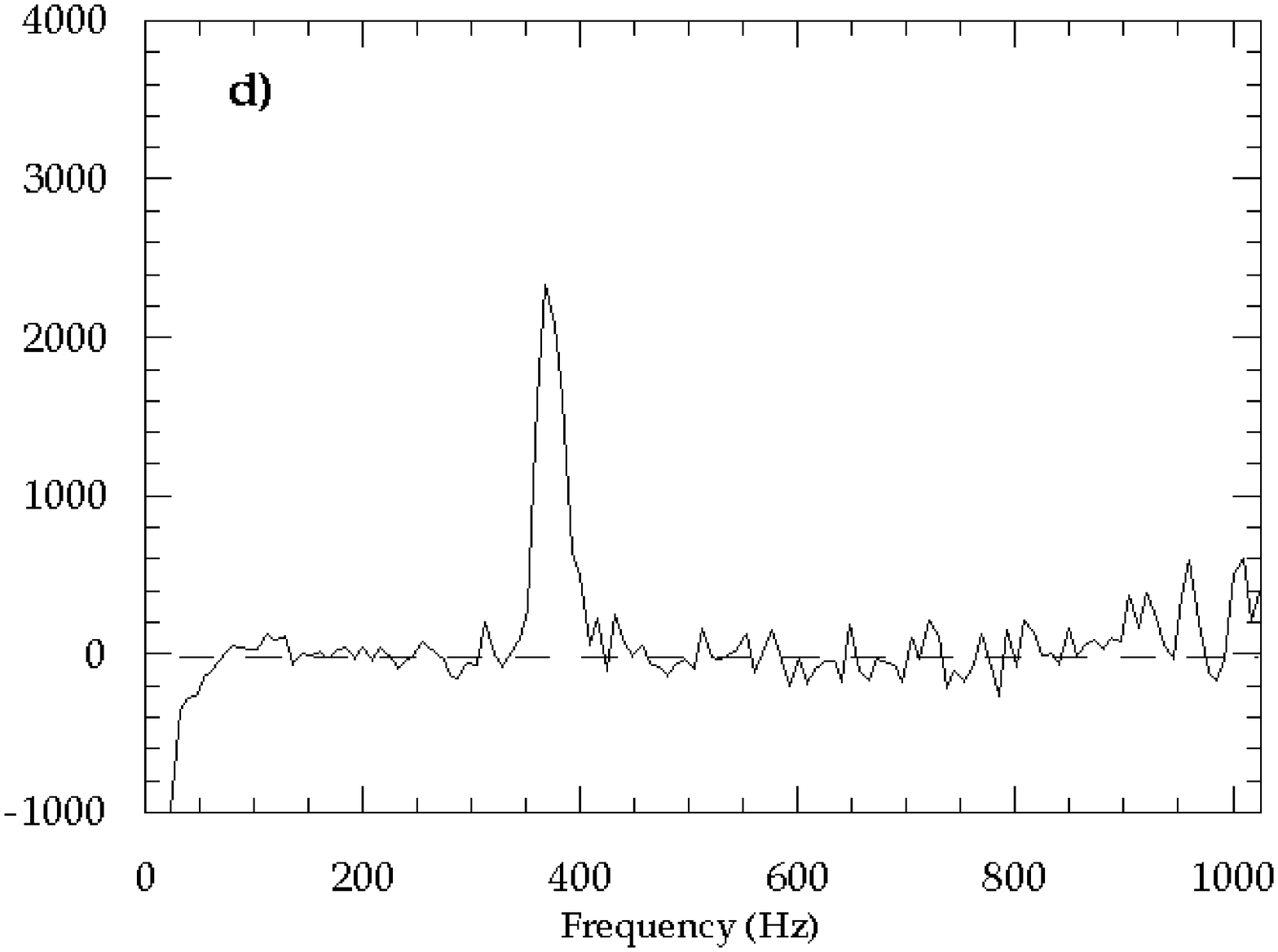}
      } } 
      
      \caption{ a) Average corrected interferograms power spectral 
      density.  b) Average processed dark current scans power 
      spectral density. c) Difference of the two to suppress the 
      bias due to detector noise. d) Difference of the two with 
      photon noise bias subtracted.  The energy peak at low frequency 
      in the processed dark signal power spectral density is due to 
      the multiplication of the dark signals by the gain function.  
      This peak causes a drop when the dark signal power spectral 
      density is subtracted from that of the fringe signal.}
         \label{fig:DSP}
   \end{center}
     \end{figure*}     
An important assumption on photon events to derive the Poisson 
statistics is that they are not correlated.  Arrival times $t_{k}$ and 
$t_{l}$ are therefore not correlated as long as $k \neq l$.  This 
important property is used to split the above expression in two terms:
\begin{displaymath}
           \left\langle \tilde{I}_{n}^{(2)}(f) \right\rangle = \nonumber
\end{displaymath}
\begin{equation}
{\left\langle {\left\langle {  \sum_{k=1}^{\tilde{N}} [g(t_{k})]^{2}
   +    2\sum_{k=1}^{\tilde{N}}\sum_{l<k} 
    g(t_{k})g(t_{l})e^{-2i \pi (t_{k}-t_{l})f}} \right\rangle}_{t_{k}} \right\rangle}_{\tilde{N}} 
\label{eq:dsp}
\end{equation}
The first average is equal to:
\begin{eqnarray}
    \left\langle 
        {\left\langle 
	     {\sum_{k=1}^{\tilde{N}} [g(t_{k})]^{2}}
         \right\rangle}_{t_{k}}
    \right\rangle_{\tilde{N}}  & = &   {\left\langle { \sum_{k=1}^{\tilde{N}} 
              {\left\langle 
	            [g(t_{k})]^{2}
               \right\rangle}_{t_{k}}
    }\right\rangle}_{\tilde{N}}  \\
    & = & {\left\langle  \sum_{k=1}^{\tilde{N}}  
              { \int_{-\infty}^{+\infty}[g(t)]^{2}\frac{i(t)}{N}dt
	 }\right\rangle}_{\tilde{N}} \\
    & = & {\left\langle \tilde{N} 
              { \int_{-\infty}^{+\infty}[g(t)]^{2}\frac{i(t)}{N}dt
         }\right\rangle}_{\tilde{N}} \\
    & = &  \int_{-\infty}^{+\infty}[g(t)]^{2}i(t)dt 
\end{eqnarray}
where I have used once again the probability density of the photon 
statistics $\frac{i(t)}{N}$. The second average of 
Eq.~(\ref{eq:dsp}) can be written as the sum of factors yielding:

\begin{displaymath}
 {\left\langle {\left\langle {  2\sum_{k=1}^{\tilde{N}}\sum_{l<k} 
    g(t_{k})g(t_{l})e^{-2i \pi (t_{k}-t_{l})f}} \right\rangle   
    }_{t_{k},t_{l}} \right\rangle}_{\tilde{N}} 
= \nonumber
\end{displaymath}
\begin{equation}
{\left\langle  {  2\sum_{k=1}^{\tilde{N}}\sum_{l<k}
   \left\langle g(t_{k})e^{-2i \pi t_{k}f}\right\rangle _{t_{k}} \left\langle 
   g(t_{l})e^{2i \pi t_{l}f}\right\rangle_{t_{l}}}  \right\rangle}_{\tilde{N}} 
\end{equation}
The averages on $t_{l}$ and $t_{k}$ can be substituted by the 
expression of Eq.~(\ref{eq:conv}) yielding:
\begin{displaymath}
 {\left\langle {\left\langle { 2\sum_{k=1}^{\tilde{N}}\sum_{l<k} 
    g(t_{k})g(t_{l})e^{-2i \pi (t_{k}-t_{l})f}} \right\rangle   }_{t_{k},t_{l}} \right\rangle}_{\tilde{N}} 
= \nonumber
\end{displaymath}
\begin{equation}
{\left\langle  
2\sum_{k=1}^{\tilde{N}}\sum_{l<k}
\frac{1}{N^{2}} |I(f) \star G(f)|^{2}
\right\rangle}_{\tilde{N}} 
\end{equation}
and:
\begin{displaymath}
{\left\langle  
2\sum_{k=1}^{\tilde{N}}\sum_{l<k}
\frac{1}{N^{2}} |I(f) \star G(f)|^{2}
\right\rangle}_{\tilde{N}} =  
\end{displaymath}
\begin{equation}
\left\langle\tilde{N}(\tilde{N}-1)\right\rangle_{\tilde{N}}\frac{1}{N^{2}}|I(f) \star G(f)|^{2} = \\
|I(f) \star G(f)|^{2}
\end{equation}
$|I(f) \star G(f)|^{2}$ is the power spectral density of the 
normalized interferogram. From the equation above I therefore derive 
an unbiased estimate of this quantity:
\begin{equation}
|I(f) \star G(f)|^{2}=\left\langle \tilde{I}_{n}^{(2)}(f) 
\right\rangle -  \int_{-\infty}^{+\infty}[g(t)]^{2}i(t)dt 
\label{eq:resultat}
\end{equation}
Since the normalized and the corrected interferograms share the same 
noise bias, the integral term of the above equation is also the bias of 
the power spectral density of the corrected interferogram.

\subsection{Comments}
The expression of the photon noise bias is very intuitive.  $i(t)$ is 
the average number of photons detected at time $t$.  Being a Poisson 
statistics, it is also the variance of the photon noise.  When the 
signal is multiplied by the gain $g(t)$, the variance at time $t$ 
becomes $[g(t)]^{2}i(t)$. Photon events at different times being 
uncorrelated, the total variance of the photon noise is therefore 
equal to the integral of the local variance. \\
Equation~(\ref{eq:resultat}) demonstrates that the noise of the corrected 
interferogram remains a white noise whose mean power spectral density 
is constant. This property is also the result of the independence of 
photon events.\\
With the above result, the computation of the unbiased estimate of 
the visibility is easy. The bias is simply obtained by co-adding the 
individual counts of the normalized interferogram. 

\section{Example of bias subtraction on real signals}
The purpose of this section is to illustrate the theoretical results 
of the previous section and to show that the assumptions on the noise 
of the  continuum and gain functions are correct. \\ \\
I have selected a series of scans of Mira observed with FLUOR in 
October 2000. The observations were carried out with a baseline long 
enough that the fringe contrast is of a few percent only. This is an 
interesting case because the source is very bright and the fringe 
contrast is small, hence the photon noise bias is relatively important 
and accounts for a large fraction of the visibility if not corrected. 
\\ \\
The raw photometric and interferometric signals are displayed on 
Fig.~\ref{fig:signauxbruts}.  The intensities are in photon counts.  
The fringes are hardly visible and their amplitude is smaller than 
that of the photometric fluctuations.  The fringes peak is visible in 
the power spectral density between 350 and 400\,Hz.  In 
Fig.~\ref{fig:continugain} are presented the gain and continuum 
functions as well as their power spectral densities on a log scale.  
The noise level of the continuum at the fringes frequency range can be 
directly compared to that of the corrected interferogram in 
Fig.~\ref{fig:signaux corriges}.  A residual photometric fluctuation 
is visible on the corrected interferogram hence the low frequency 
component.  In the data reduction procedure, the detector dark current 
signals undergo the same correction process as the source signals.  
{In case the dark currents would combine both detector dark signal and 
sky background if measured with a chopping technique, the dark signals 
would contain both background photon noise and detector noise.  But 
both noises can be treated as additive noise, in particular the 
background photon noise does not need to be processed like the source 
photon noise.  In the case of the FLUOR signals, background photon 
noise is totally negligible and the dark signals only contain the 
detector contribution.} In Fig.~\ref{fig:DSP} I have represented the 
average of power spectral densities of both the corrected detector 
dark current and corrected interferograms to reduce the noise on the 
power spectral densities.  On the same figure, the difference of the 
two is plotted. At this stage, the bias due to source photon 
noise is not removed to make its magnitude obvious to the reader. The 
fringe squared visibility being obtained by integrating the power 
spectral density at the fringe frequency, the non-zero mean level 
under the fringe peak causes the bias.  This is the bias due to source 
photon noise.  This graph also shows an increasing residual noise at 
high frequency due to detector instabilities.  The value of the bias 
has been computed with the method presented in this paper and 
subtracted as illustrated by the last graph of this figure.  The power 
spectral density background level is now equal to zero in average 
showing that the bias has been correctly subtracted.

\section{Conclusion}
A method to subtract the photon noise bias from visibility data 
acquired with a single-mode interferometer has been presented. An 
analytical expression for the bias can be established under verified 
assumptions. Other non-analytical methods based on the fit of 
the average level of power spectral densities may suffer from 
confusion with fringe signal and from detector instabilities and the 
analytical method should be preferred.

\begin{acknowledgement}
    I wish to thank P. Bord\'e for his careful reading of the paper 
    and for his precious comments that helped me write the final version. 
\end{acknowledgement}


\begin{thebibliography}{}    
\bibitem[Coud\'e du Foresto et al. (1997)]{foresto1997} Coud\'e du Foresto, 
V., Ridgway, S.T., Mariotti, J.-M., 1997, A\&ASS, 121, 379
\bibitem[Goodman (1985)]{goodman1985} Goodman, J., 1985, Statistical 
Optics, J. Wiley \& Sons Ed., 43
\bibitem[Perrin (2002)]{perrin2002} Perrin, G., 2002, submitted
\bibitem[Tango \& Twiss (1980)]{tango1980} Tango, W.J., Twiss, R.Q., 
1980, Progress in Optics 17, E. Wolf Ed., 239
\bibitem[Thi\'ebaut (1994)]{thiebault1994} Thi\'ebaut, \'E., 1994, PhD 
thesis, Universit\'e Paris VII 
dissertation
    


  \end{thebibliography}
\end{document}